# Title: Agroseismology: unraveling the impact of farming practices on soil hydrodynamics

**Authors:** Qibin Shi (*,[1,2]), David R. Montgomery[1], Abigail L.S. Swann[3,4], Nicoleta C. Cristea[5], Ethan Williams[1,6], Nan You[7], Joe Collins[8,9], Ana Prada Barrio[8], Simon Jeffery[8,9], Paula A. Misiewicz[8], Tarje Nissen-Meyer[9,10], and Marine A. Denolle (*,[1])

**Affiliations:**

[1] Department of Earth and Space Sciences, University of Washington; Seattle, USA.

[2] Department of Earth, Environmental and Planetary Sciences, Rice University; Houston, USA.

[3] Department of Atmospheric and Climate Science, University of Washington; Seattle, USA.

[4] Department of Biology, University of Washington; Seattle, USA.

[5] Department of Civil & Environmental Engineering, University of Washington; Seattle, USA.

[6] Department of Earth and Planetary Sciences, University of California; Santa Cruz, USA.

[7] Department of Earth, Atmospheric, and Planetary Sciences, Purdue University; West Lafayette, USA.

[8] Agriculture and Environment Department, Harper Adams University; Newport, UK.

[9] Earth Rover Program; London, UK.

[10] Department of Mathematics and Statistics, University of Exeter; Exeter, UK.

*Corresponding author. Email: qshi@rice.edu, mdenolle@uw.edu

**Abstract:**

**Farmed landscapes provide a natural laboratory to test how management reshapes near-surface hydrodynamics. Combining distributed acoustic sensing with physics-based hydromechanical modeling, we tracked minute-resolution, meter-scale changes across experimental fields with controlled tillage and compaction histories. We find that dynamic capillary effects—rate-dependent suction stresses during wetting and drying—govern transient stiffness and moisture redistribution in disturbed soils, producing sharp post-rain velocity drops from near-surface saturation and large hysteretic velocity rebounds driven by evapotranspiration. By pairing a seismic rainfall proxy with a drainage closure, we invert velocity changes to estimate evapotranspiration, revealing how disturbance alters flux partitioning and storage. These results establish agroseismology as a non-invasive, extendable tool to uncover soil hydromechanics, explain why conventional farming intensifies variability, and provide new constraints for Earth system models, land management, and hazard resilience.**

## Introduction

Soils form our planet's thin yet vital skin—a dynamic interface where the atmosphere meets solid Earth, governing the exchange of water, carbon, and energy while underpinning agricultural productivity (50% of global habitable land use), climate regulation, and landscape stability under both natural and engineered conditions. At the heart of these functions lies the soil's porous architecture, which controls water retention and movement—balancing soil moisture evaporation, plant transpiration, and water infiltration, thereby shaping surface and subsurface hydrological responses (*1*). Natural structures that develop over decades through intertwined biological and geomorphological processes can be disrupted by modern land management with far-reaching consequences—changing subsurface water dynamics central to soil ecosystems (*2*), interrupting evapotranspiration and energy fluxes essential for land surface modeling and climate dynamics (*3*), modifying mechanical properties relevant to geotechnical and earthquake engineering (*4*), and undermining moisture availability and resilience in agriculture (*5*). Nevertheless, most land surface and hydrological models treat soil as a static medium, overlooking its time-varying properties (*6*) and effects from repeated mechanical disturbance.

Current approaches to measuring soil's hydrological properties include both ground-based methods using point measurements with *in-situ* sensors (e.g., time-domain reflectometry, capacitance, impedance sensors, thermal probes (*6*)) and non-invasive sensing (e.g., cosmic-ray neutron sensing, global navigation satellite systems reflectometry, gamma-ray monitoring, ground-penetrating radar (*7*)). However, spatial heterogeneity makes it difficult to scale ground-based measurements to a large spatial extent. Satellite and airborne sensing (e.g., SMAP, SMOS) solve this with regional and global coverage (*8*), primarily of moisture, but spatial resolution is often coarse (tens of kilometers). Higher-resolution products from synthetic aperture radar carry sub-kilometer spatial resolution (*9*, *10*) but suffer from low temporal resolution (days to months) and are sensitive to vegetation, surface roughness, and atmospheric noise, limiting their ability to resolve fine-scale and short-term soil moisture dynamics. Capturing rapid changes in soil water content driven by evapotranspiration (*ET*), precipitation, and drainage under heterogeneous soil conditions and understanding the impact of anthropogenic disturbance remain critical challenges essential for guiding agricultural practices, mitigating soil degradation, and optimizing water use efficiency.

Structural changes in soils significantly influence stiffness and seismic wave speeds (*11*). Seismological methods have recently emerged as a powerful tool to repurpose earthquake sensors for continuous, spatially distributed monitoring of subsurface hydromechanical properties from changes in seismic velocities ($dv/v$) that have been shown to vary with groundwater and soil moisture levels (*12–16*). Modeled as responses to pore pressure changes due to rainfall diffusion (*12*, *14*) under the assumption of nonlinear elasticity in saturated media (*17*, *18*), such correlations have permitted empirical assessments of droughts (*13*, *16*, *19*, *20*). Yet, to date, the approach remains anecdotal and requires site-specific tuning to explain observations (e.g., (*18*)), highlighting the need for better physical models of $dv/v$ in relation to soil moisture and structure.

Static capillary forces in partially saturated soils play an important role in soil's stiffness (*11*) and distributed acoustic sensing (DAS) transforms standard fiber-optic cables into dense seismic arrays with meter-scale spatial resolution, sub-minute temporal resolution, and broad frequency sensitivity, allowing detection of shear-wave velocity changes linked to near-surface moisture variability (*21*). The utilization of existing fiber-optic cables would enable continuous monitoring of soil hydrodynamics over tens of kilometers at meter-scale resolution, uniquely addressing the scale gap between traditional point sensors and remote sensing. Here, we leverage DAS and hydromechanical modeling to understand the seismic signature of soils in an experimental farm where long-term tillage and compaction treatments have consistently altered soil structure. This long-term experimental farm provides a rare opportunity to test how controlled disturbance regimes reshape soil hydromechanics systematically. By integrating DAS with physically grounded models, we unravel how dynamic soil properties mediate the coupling between water fluxes and mechanical stability, with fundamental implications for agriculture, land surface modeling, meteorological forecasting, and geotechnical risk analysis.

## Data acquisition

From March 17 to 19, 2023, we conducted a DAS experiment on fields at Harper Adams University, UK (Lat. 52.782, Long. -2.428) across 27 ~4-meter-wide plots, that had each received consistently a unique combination of tillage and compaction since 2011 (Fig. 1). Tillage was applied to three depths—no tillage, 10 cm, and 25 cm—while compaction was induced by varying tire pressure (and its subsequent soil-surface contact) at two pressure levels—70 kPa for both front and rear tires, and 120 kPa for front and 150 kPa for rear tires. Soil porosity measurements confirmed significant spatial variability across the plots (Fig. 1B). We deployed a fiber-optic cable perpendicular to boundaries between the plots, in a finger-wide trench directly into the bare soil to an average depth of ~2 cm to ensure good soil coupling (Fig. 1E). Using a Sintela Onyx 1.0 DAS interrogation unit, we recorded continuous ambient seismic data for 40 hours at 2 kHz sampling rate and 3.19 m interval between channels, which were used to track different plots. Spectral analysis revealed one or two spectral peaks: typically, a fundamental mode around 15–25 Hz and an overtone near 25–50 Hz (fig. S1). These are consistent with resonance from soft topsoils overlaying stiffer subsoils. These peak frequencies are functions of topsoil thickness and depth-integrated shear-wave velocity ($v_S$) along the cable ((*22*), Fig. 1D, and fig. S1). Episodes of broadband seismic energy, including notable energy above 80 Hz, corresponded with rainfall, consistent with prior findings that precipitation rate is the dominant control at these frequencies (*23–25*), with additional influences from raindrop velocity (*25*) and droplet size distribution (*23*). Simultaneously, we collected meteorological data from a nearby weather station, including air temperature, humidity, and hourly precipitation rate. Using this precipitation data, we validated and calibrated the DAS-derived power spectral density (PSD) in the 80-140 Hz band as a 1-minute resolution proxy rain gauge (Fig. 1F and fig. S2). Additional weather information, such as cloud cover and wind speed, was collected from a regional station ((*22*), fig. S3). Overall, daily rainfall ranged between 5-10 mm, temperature between 5-15°C, humidity between 60-95% and wind speed (at 10-meter height) between 1-10 m/s.

We detected temporal changes in seismic velocity ($dv/v$), which were primarily sensitive to changes in $v_S$ (*26*), using ambient noise auto-correlation functions (ACF; (*27*)) derived from DAS. ACFs were stable over 15-s windows and showed coherent evolution between 15–60 Hz across plots (Fig. 1G, figs. S4 and S5). A three-pass denoising and polynomial fitting routine improved robustness, especially during rapid $dv/v$ fluctuations (e.g., after rainfall; fig. S6). The resulting $dv/v$ time series (e.g., Channel 18, hereafter as Ch 18; Fig.1G) showed clear negative correlation with soil water content: precipitation events (P1-P4) triggered sharp decreases in $dv/v$, followed by gradual recovery during drying. The $dv/v$ response was spatially variable—about 60% reduction in $dv/v$ in response to drying was shown at some plots (e.g., near Ch 18; Fig. 1G), which marked the highest variability, while much more minor changes on other plots were observed (e.g., Ch 33; fig. S5). We identified *precipitation*, *evapotranspiration*, and *drainage* as the three primary processes driving soil moisture dynamics, similar to (*28*). To move beyond empirical observations and explain why soils under different treatments responded so differently, we developed hydrological and hydromechanical models linking seismic velocity to soil water saturation and pore-scale structure.

**Physical models**

To interpret the observed seismic velocity changes, we built models that link soil water saturation to shear-wave velocity by incorporating hysteretic, capillarity-controlled effective stress into Hertz-Mindlin contact theory (*29*). Soil water saturation, $S_w$, is defined as the volumetric fraction of water in pore space (*11, 21*). Our framework combines two elements: (i) a hydrological model that captures how precipitation, evapotranspiration, and drainage control water saturation, and (ii) a lithological model that translates changes in saturation into stiffness and seismic velocity.

The lithological model ***L(.)*** accounts for the impact of varying $S_w$ on shear-wave velocity $v_S$. We start with $v_S = \sqrt{\mu_{eff}/\rho}$, where both effective shear modulus $\mu_{eff}$ and density $\rho$ depend on $S_w$. $\mu_{eff}$ derives from Hertz-Mindlin theory (*28*), incorporating grain contact mechanics and $S_w$-dependent effective pressure $P_e$ (*22*). $P_e$ includes a capillary component that provides suction pressure, counteracted by gravity, enhancing soil frame rigidity. While static suction stiffens soils during drying, only dynamic capillarity—stresses that depend on wetting or drying rates (*30*)—explains the observed short-term hysteresis in seismic responses (fig S7). In the framework of (*31*), the dynamic capillary effect is related to the finite redistribution time of water flow in the pore networks, thus dependent on the hydraulic conductivity and intensities of meteorological forcing.

To link soil stiffness with meteorological forcing, we coupled this lithological model to a hydrological model ***H(.)*** that tracks the balance between precipitation, drainage, and evapotranspiration, following (*27*): $\dot{S_w}(t) = P(t) - Q(t) - ET(t)$ , where $\dot{S_w}$ is the time derivative of water saturation, $P$ is precipitation, $Q$ is vertical drainage, and $ET$ is evapotranspiration. We parameterized the drainage with an exponential decay (~70-hour timescale) on the precipitation input, similar to a baseflow model (*32*), and estimated $ET$ from weather station data (Fig. 1F) using the Penman-Monteith equation (*33*) and adjusted for treatment effects (*22*).

We tested the combined model $v_S = L[H(P, Q, ET)]$ against observed velocities $v^{obs} = v_{S,ref} * (1 + dv/v)$ using the modelled $v_S$ immediately after rainfall event P4 as $v_{S,ref}$. We compared clay- and sand-type soil textures to represent endmember grain contacts: with more cohesive grain-to-grain contacts, the clay-type texture serves as the upper bound of soil stiffness, while the sand-type texture represents the lower bound.

We also tested whether the various capillary models (their absence, or whether they are static or dynamic) best explain hydrological and lithological variabilities at different channels. We illustrate two contrasting responses to $ET$ (March 18, 12:00; Fig. 2), which either exhibit strong temporal fluctuations at Ch 18 or do not show any at Ch 33. The substantial $v_S$ rise of 60% at Ch 18 was reproduced by dynamic capillary pressure and a rapid decrease in $S_w$ (as much as 50% in a single afternoon) driven by $ET$ (Fig. 2B), whereas static capillary effects explained only about half. In contrast, Ch 33 showed low temporal variability (Fig. 2C) despite identical meteorological forcing. Other channels (e.g., Ch 44; fig. S8) showed intermediate variability, consistent with static capillary effects under high hydrological variability. These stark differences, despite close spatial proximity, point to soil structure—shaped by tillage and compaction—as the dominant control, motivating a direct analysis of hydrological variability across soil treatments.

Based on our lithological model, the moisture perturbation due to a given hydrological process (e.g., drainage) is proportional to the relative change in shear-wave velocity. This proportionality reflects variability in pore-fluid pressure, with additional contributions from capillary effects ((*22*) Section 6; fig. S9). The hydromechanical response is strongly modulated by structural alterations from tillage and compaction. We find four distinct styles of lithological responses in the DAS analysis that correspond to different combinations of tillage depth and compaction pressure (Fig. 3A). All styles exhibit varying degrees of evaporation and drainage, shaped by specific near-surface regimes.

The primary contrast lies between the response style showing nearly zero saturation perturbation over two days (Fig. 3A; green) and styles with significant perturbations (Fig. 3A; red, orange, yellow). The former corresponds to little or no tillage, representing an unsaturated flow regime where water from precipitation or irrigation rapidly infiltrates into the vadose zone via a well-connected pore network. The latter corresponds to deep tillage, high compaction, or a combination of both, which disrupts pore connectivity, impedes drainage, and results in transient near-surface saturation. Common causes of such disruption include a compacted "plow pan" layer at tillage depth, which impedes vertical flow, and matrix collapse and macropore closure from high compaction (*34*). Evidence from the correlation between air humidity and soil saturation beneath Ch 18 (Fig. 2A) indicates that moisture remained near the surface without penetrating to depth— a hallmark of impaired hydraulic connectivity. These alterations reduce permeability, limit infiltration depth, heighten sensitivity to surface evaporation, and promote runoff and thereby erosion. In undisturbed soils, pore networks enable infiltration to the root zone and beyond. In disturbed soils with reduced infiltration, near-surface saturation develops, increasing runoff and evaporation, and thus delivering less moisture to the root zone (Fig. 4).

Three different styles of saturated response reflected treatment intensity. Deep tillage and high compaction both produced pronounced 5-hour evaporation and prolonged drainage, while the combination of moderate tillage and compaction showed little evaporation (Fig. 3D), shorter drainage durations, and slightly faster post-rain drainage (Fig. 3C). Evaporation rates also differed between extreme tillage and extreme compaction due to their distinct impacts on pore structure. High surface loading pressure from traffic (Fig. 3A, dotted contours) compacts near-surface soil, altering the vertical distribution of permeability and aggregate connectivity. Under shallow tillage and high compaction, micropores dominate water flow, promoting capillary rise and prolonged "Stage I" evaporation. In contrast, under deep tillage and low compaction, near-surface pore networks are disrupted, producing less efficient "Stage II" evaporation. In other words, the contribution of capillary effects to the velocity change is controlled by the distribution of water flowing through micropores and macropores. Whether water flows preferentially through micro or macropores will control the style of capillary effects (i.e., dynamic or static); a combination of these effects may take disturbed soils.

To characterize the mechanical disturbance of each plot, we parameterized tillage depth, $T$, in cm, and compaction (tire pressure), $C$, in kPa, and defined a disturbance index $DI$. We chose a product and power-law functional form $DI = T^n C^m$, normalized $DI$ between 0 and 1, and fit the cumulative drainage $\delta(dv/v)$ for the exponents via grid search. We found a best fit of n = 0.22 and m = 0.96 (fig. S10) and show the spatial variation of these correlations in Fig. 3B. Our results confirm a strong impact of mechanical disturbance (Fig. 3A) on hydrological and hydromechanical variability, confirming previous reports (*34*). We model $v_S$ across DAS channels based on $DI$, which aligned with observed patterns in space and time (Fig. S11), demonstrating the compounding effects of tillage and compaction on soil structure.

High-*DI* soils (Fig. 3A; red, orange, yellow) have decreased pore connectivity, heightened capillary suction, and prolonged water redistribution. These dynamic capillary effects (*31*) amplified the soil's response to atmospheric forcing, driving greater fluctuations in inter-particle stress as well as in shallow moisture content during cycles of rainfall and evaporation. Over time, these effects can contribute to pore-scale mechanical fatigue, likely weakening soil aggregates. Low-*DI* soils (Fig. 3A; green) maintain well-connected pore networks that promote deep infiltration and capillary buffering, thereby stabilizing shallow soil moisture across wetting and drying periods.

These hydraulic and hydromechanical effects are rooted in the mechanical disturbance of the soil structure. The sensitivity of seismic velocity to water saturation enables us to observe, in situ and in real time, the coupling between pore connectivity, moisture retention, and hydrological redistribution rates. By capturing both rapid hydrological fluxes and slow mechanical recovery, our approach bridges the physical models proposed by hydrologists with the dynamic behavior observed in the field—revealing how soil structures profoundly influence hydrological function. These insights carry significant implications for guiding sustainable agriculture and water management, as well as for improving models of land surface fluxes, carbon cycling, greenhouse gas fluxes, and geotechnical risk across disturbed and undisturbed landscapes.

## Discussions

These discoveries directly translate into consequences for agriculture, where soil disturbance practices alter both hydraulic function and resilience. Our findings show that mechanical disturbance—through repeated tillage and compaction—degraded both the hydraulic and hydromechanical stability of soils at spatial and temporal scales, which has not been possible to explore previously. Through reducing effective porosity and effectively reducing near-surface hydraulic conductivity, tillage- and compaction-related disturbance traps moisture near the surface, impedes infiltration, and amplifies dynamic capillary stresses that stiffen the soil during drying and may contribute to mechanical fatigue that weakens aggregates and accelerates erosion (35). Seismic monitoring uniquely captures these processes in near-real time, validating long-standing hypotheses in soil science about pore connectivity, soil water retention, and loss of resilience under conventional practices (5, 36). Such structural changes compromise subsurface water storage critical for plant health, increasing erosion, runoff, and evaporative losses. Our study provides mechanistic evidence that low-disturbance farming practices exhibit more stable subsurface moisture dynamics. Specifically, our results confirm that high-disturbance farming practices amplify hydrologic variability, whereas low-disturbance farming practices maintain soil structure that buffers soil moisture volatility. By enabling high-resolution, non-invasive monitoring of soil water and structure in situ, seismology opens new possibilities for managing soils as climate-adaptive infrastructure. Preserving and improving effective porosity and moisture buffering supports agricultural productivity, mitigates carbon loss, and reduces agricultural greenhouse gas emissions (37). Real-time quantification of soil moisture at the landscape scale has the potential to greatly improve meteorological forecasting (38).

Beyond the farm scale, these soil–water interactions shape how land–atmosphere exchanges are represented in climate models, with direct implications for constraining uncertainty and improving climate resilience. Recent analyses indicate that soil processes and structure exert some of the strongest controls on land-atmosphere interactions (39), and incorporating structural effects could substantially improve predictions of surface–subsurface fluxes and plant–soil interactions (40, 41). Yet, soil disturbances remain underrepresented in current hydrologic models (42), and their impacts on the land surface dynamics and groundwater recharge are recognized as major unresolved problems in hydrology (43). Our results provide a pathway to address this gap by (i) tracking a prognostic structural state that evolves with land management and wetting–drying cycles, and (ii) incorporating rate-dependent capillarity terms to capture the short-term stiffness and moisture fluctuations, and, most importantly, to estimate soil water balance for re-partitioning

*ET*, runoff, and recharge (e.g., Fig S12). This approach offers a high-resolution alternative to traditional energy-based *ET* estimates, thereby reducing uncertainty in climate-relevant land–atmosphere coupling (*44*).

At coarser scales (1-10 km), Earth system models typically parametrize soil hydrological processes using emergent, aggregated water fluxes via statistical or physics-based approaches (*45*) and address the sub-grid heterogeneities using soil tiles (*42*). Given the critical influence of soil processes on the climate system, proper calibration of these parameterizations is essential. DAS emerges as a promising technology, providing meter-scale resolution across tens of kilometer-scale, bridging plot-scale processes with landscape- and watershed-scale dynamics. This scalability is particularly valuable in intensively managed agricultural regions and lowlands. Our results highlight the potential for fiber-optic sensing to further constrain water and energy fluxes at the soil–atmosphere interface, informing land management strategies for climate adaptation.

The hydromechanical variability we observe also extends beyond agriculture and climate into geotechnical engineering, where moisture-dependent soil stiffness impacts hazard and infrastructure stability. This study shows that the hydromechanical impacts of tillage and compaction fundamentally alter the soil's seismic behavior. While near-surface $v_S$ has long been central to site response models (e.g., (*46*)), our seismic measurements reveal that time-dependent changes—driven by soil moisture and capillary-induced effective stress—challenge the assumption of a stationary shallow layer, consistent with growing evidence that groundwater fluctuations and seasonal variability can influence site response and ground failure (*47*, *48*). Additionally, earthquake-triggered liquefaction, which has traditionally been considered a phenomenon limited to fully saturated soils, has been recently found to occur in partially saturated soils (e.g., (*49*)) once a critical percolation threshold is exceeded (~70% saturation). These processes are especially variable in mechanically disturbed soils, where heterogeneity in pore structure modifies the local stress path. As climate variability and agricultural intensification reshape soil hydrodynamics and depreciate earthquake resilience, accounting for these effects will be essential for predicting ground failure, guiding mitigation, and designing seismically and hydrologically resilient infrastructure. One emerging ground improvement strategy, beyond soil desaturation (*49*) microbially induced calcite precipitation, depends critically on managing subsurface fluid flow and mechanical evolution through biological processes (*50*), which naturally occur in undisturbed and healthy soils. Our fiber-optic seismic method offers a powerful means of monitoring such treatment in situ, providing real-time feedback on stiffness development and long-term post-treatment surveillance.

## Conclusions

Our study shows that fundamental physical mechanisms—dynamic capillary effects in managed soils—govern how water redistributes under atmospheric forcing and alters mechanical stiffness in real time. This mechanism reveals that soil structure, shaped by decades of land use or minutes of rainfall, actively modulates evapotranspiration, infiltration, and seismic response, with cascading consequences for agriculture, climate, and hazard resilience. These insights recast the shallow subsurface not as a passive boundary layer, but as a dynamic system whose behavior underpins land–atmosphere coupling and the stability of infrastructure. By capturing these processes with fiber-optic sensing and physics-informed models, we offer not only a mechanistic advance but also a technological blueprint for a new class of Earth system observatories. As we move toward real-time, multiscale digital twins of the soils, our approach establishes the necessary foundation—linking soil physics, hydrology, and seismology to anticipate how soils evolve under climate extremes and anthropogenic stress.

**Acknowledgments:**

The authors thank the University of Washington FiberLab for instrumental support, and Dominik Gräff and Kuan-Fu Feng for their helpful discussions.

**Funding:**

The Pan Family Fund

Murdock Charitable Trust

UW College of Env Seed Fund

Packard Foundation

NERC Cross-disciplinary research capability grant

**Author contributions:**

Conceptualization: MD, TN, SJ, QS, DM, NC

Methodology:  QS, NY, MD

Investigation:  QS, MD, NY, AS, EW, NC

Data curation: QS, EW, APB, JC, PM

Visualization: QS

Formal analysis: QS, MD

Validation: QS, MD, PM

Software: QS, NY

Funding acquisition: MD, TN, SJ, AS, DM

Project administration:  MD

Supervision:  MD, DM, AS, NC, SJ

Writing – original draft: QS, MD

Writing – review & editing:  QS, MD, DM, AS, NC, TN, SJ, PM, EW, NY, APB, JC

**Competing interests:** The Authors declare that they have no competing interests.

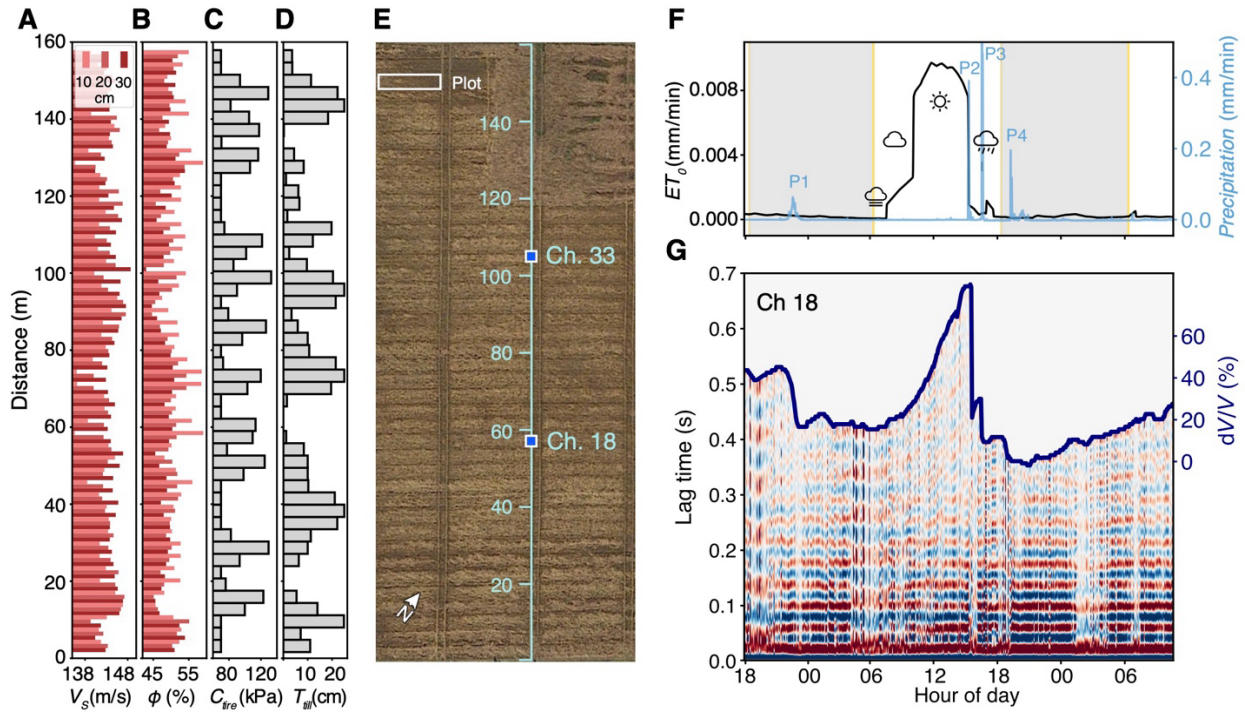

**Fig. 1. Spatiotemporal variations of soil properties, treatments and hydrological forcing.**

(**A**) Predicted shear-wave velocity ($v_S$) assuming saturated soil, shown at different depths and locations along the cable. (**B**) Soil porosity at different depths and locations. (**C**) Degree of compaction from tire pressure ($C_{tire}$) applied during traffic, interpolated onto DAS channels. (**D**) Tilling depth ($T_{till}$) interpolated onto DAS channels. (**E**) Map of the research farm with the fiber-optic cable indicated in light blue. (**F**) Precipitation rate highlighting four main rainfall events (P1-P4), denoted by high-frequency DAS PSD, aligned with reference evapotranspiration ($ET_0$) from net radiation and hourly weather conditions (symbols). (**G**) Temporal changes in velocity over 40 hours at 58 m from the south end, together with the stretched auto-correlation functions.

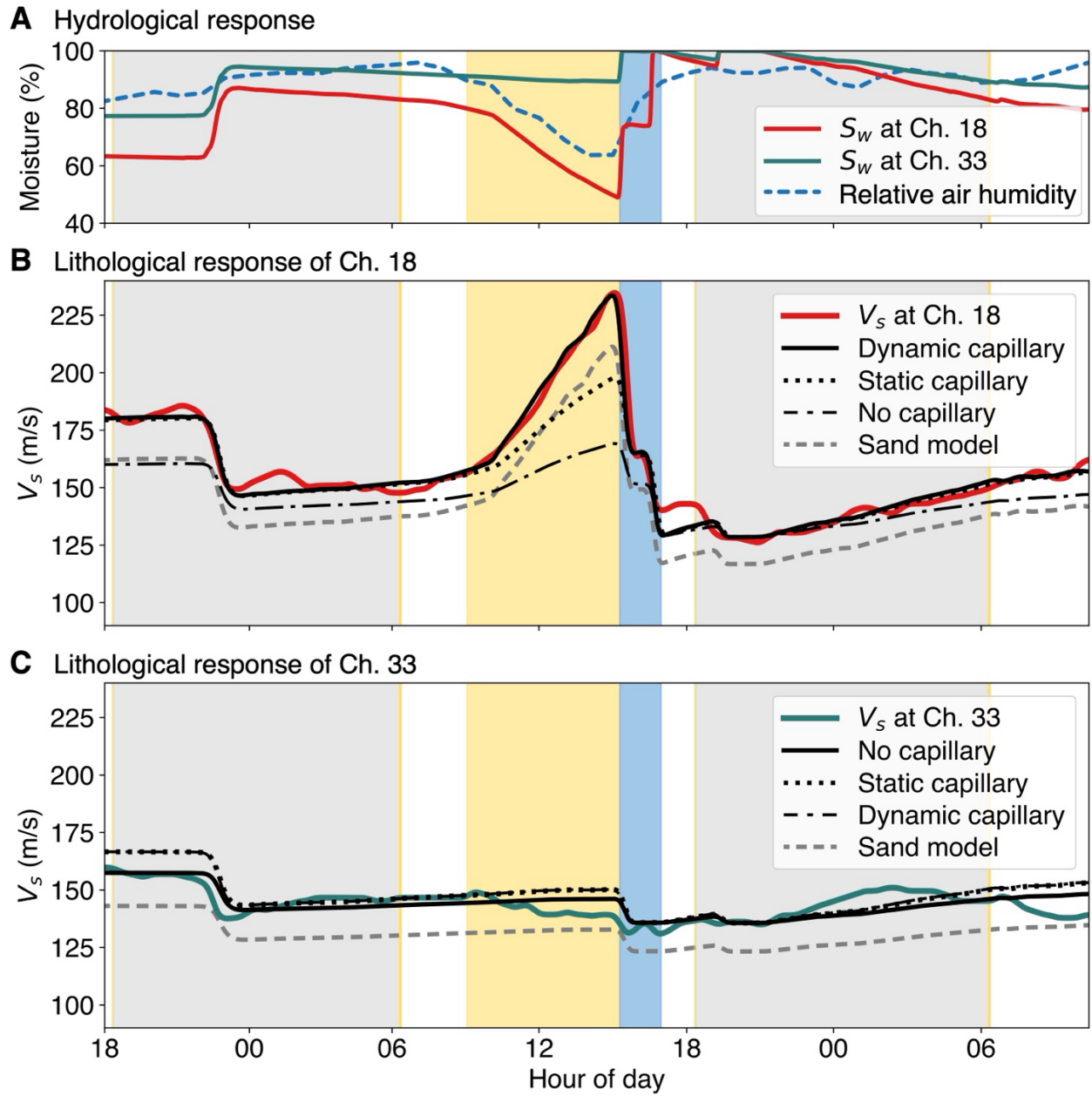

**Fig. 2. Hydrological and lithological responses.**

(**A**) Water saturation of Ch 18 and Ch 33 during hydrological events, compared with air relative humidity. (**B**) Observed and modeled shear-wave velocity in response to lithological changes near the Ch 18. (**C**) Same as (B), but for Ch 33.

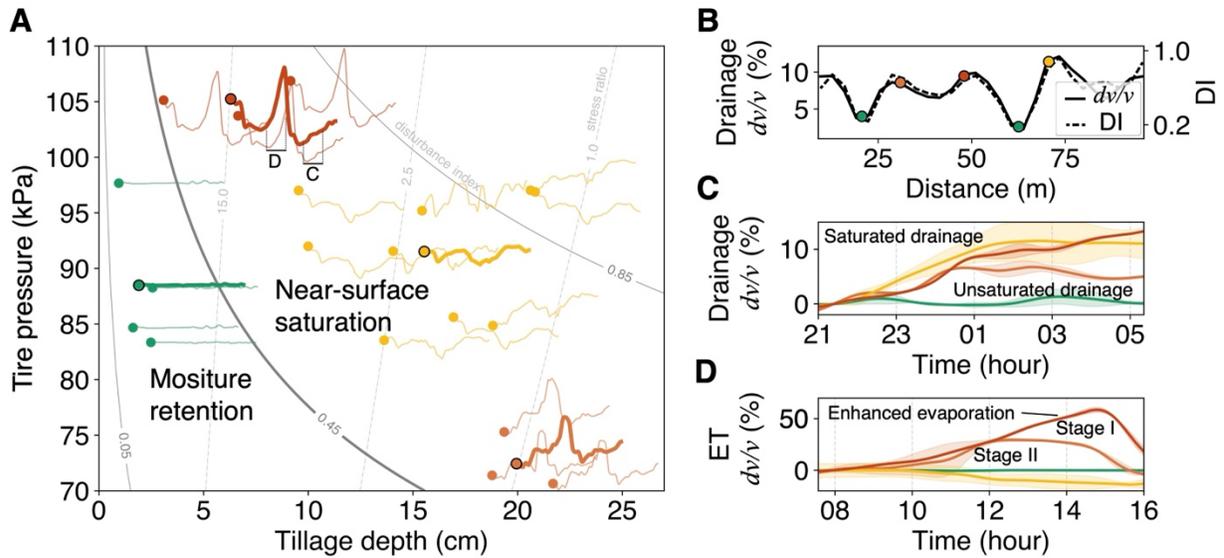

**Fig. 3. Spatially variability of mechanical disturbance.**

(**A**) Spatial variation of hydromechanical responses in relation to tillage depth and degree of compaction measured by tire pressure. Four groups of $dv/v$ curves, each negatively correlated with water saturation, are located using their first data point and color-coded in a traffic-light sequence according to relative hydrological response levels. Contour lines represent the disturbance index (DI) and the relative impact of surface compaction on tillage depth. (**B**) Disturbance Index (dashed) and total velocity change (solid) along the cable during the drainage-dominated period marked "C" in (A). (**C**) Velocity changes governed by saturated versus unsaturated drainage during the interval "C" in (A). (**D**) Velocity changes governed by evaporations during interval "D" in (A).

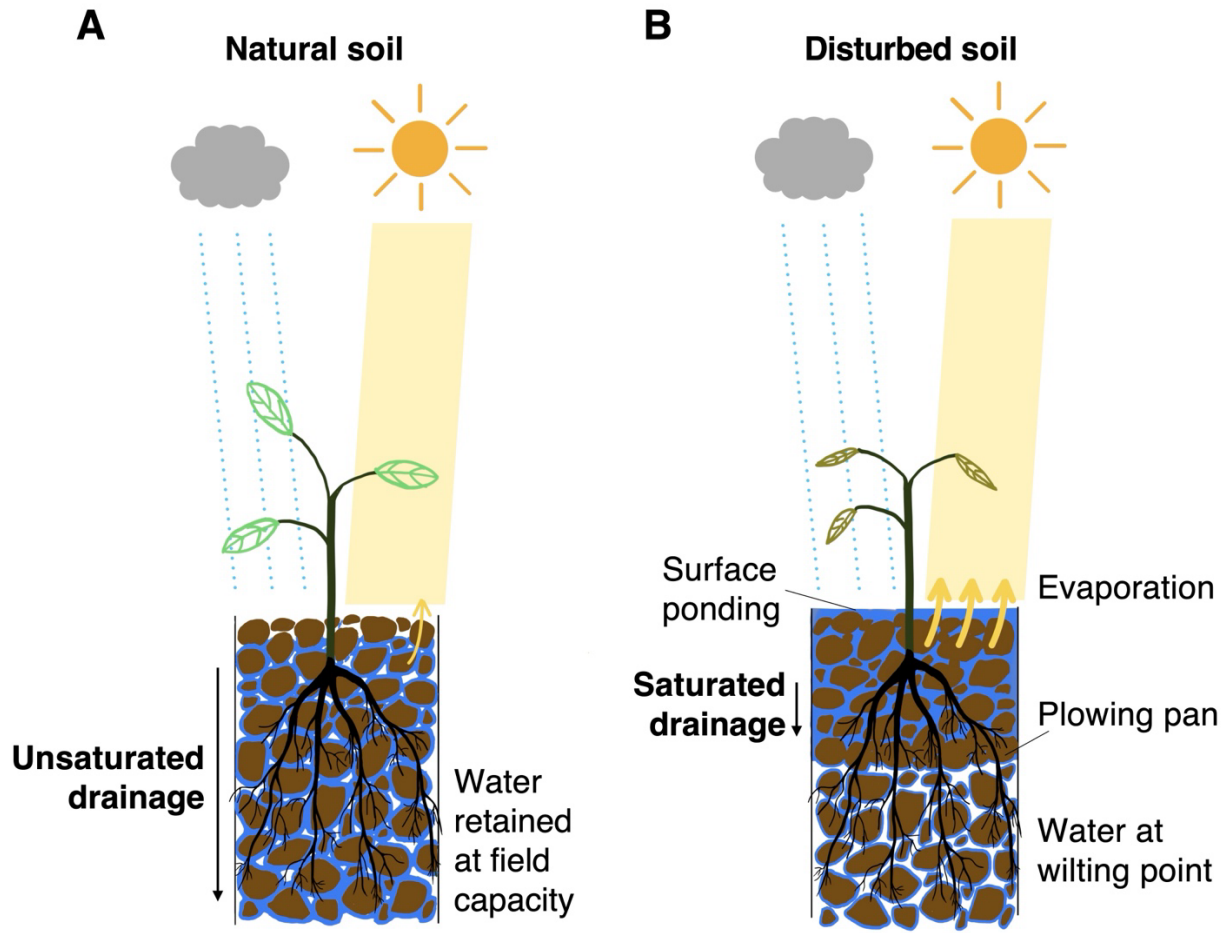

**Fig. 4. Conceptual porous model of natural and disturbed soils.**

(**A**) In natural, undisturbed soil, unsaturated drainage can retain moisture at the field capacity, enabling root water uptake. (**B**) In disturbed soil, structural alternation leads to near-surface saturation, dry below the tillage depth, and enhanced evaporation.

**Supplementary Materials**

1. Weather data

We use two sources of weather data in our analysis. The first source is from a nearby weather station in Newport, Shropshire, UK. It provides data at 30-minute intervals, including air temperature, weather conditions, wind speed, and humidity. This data was last accessed on 2024/09/24 at https://www.timeanddate.com/weather/@7296017/historic?month=3&year=2023. The data is shown in fig. S3.

The second source comes from Harper Adams University, UK (Lat. 52.778, Long. -2.428). At the time of collection, this dataset had a 1-hour resolution and was converted into a tabular format. It includes soil temperature measurements at depths of 10, 30, and 100 cm, cumulative rainfall gauge readings, and air humidity. Hourly rainfall is derived by differencing the cumulative gauge readings, with a daily calibration applied to the first data point of each day. The data that we used is shown in fig. S3.

2. Soil and plot property data

The long-term traffic and tillage project was established in 2011 on sandy loam soil, at Harper Adams University, UK. The experiment consists of 4 replicated blocks, set up in a randomized block design. The three traffic management systems were: STP (standard inflation pressure tires, front: 120 kPa, rear: 150 kPa), LTP (low inflation pressure tires, front and rear: 70 kPa) and CTF (controlled traffic farming, front and rear: 70 kPa). The three tillage systems were: Deep (25 cm), Shallow (10 cm) and Zero tillage. Each plot is 3.6-4.5 m wide and 85 m long.

The field was tilled on March 7, 2023, and drilled on March 8, 2023, with spring oats. The DAS experiment was conducted from March 17 to 19, 2023. Bulk density samples were collected on September 18, 2023, in all plots, to 30 cm. Samples were separated by three depth increments (0-10, 20-10 and 20-30 cm), resulting in 432 samples (36 points x 3 depths x 4 times). Stones were manually removed from the soil and weighed before drying. Samples were collected in the middle of the plot (non-trafficked crop area) for CTF, and the trafficked area (permanent wheel way) for STP and LTP systems. The drainage map was produced by Hawkins Drainage Systems (Ellesmere, UK) when the drains were installed in 2011.

3. Resonance Frequency

From March 17 to 19, 2023, we conducted a distributed acoustic sensing (DAS) experiment on fields at Harper Adams University and collected continuous DAS data as consecutive 1-minute HDF5 files with 2 kHz sampling rate and 3.19-m channel spacing. In the pre-processing step, we downsampled the time-series data to 200 Hz and selected only 51 out of 200 channels that are well-coupled with the soil.

We estimated spectrograms using the short-time Fourier transform, with a window length of 81.92 seconds, 10.24-second overlap, and frequency content up to 50 Hz. Across all channels, the spectrograms consistently reveal one or two resonance frequencies. In some channels, these frequencies vary over time, while in others, they remain stable. Overall, the analysis identifies a fundamental mode typically in the range of 15–25 Hz, and an overtone between 25–50 Hz. Three examples of spectrograms at Ch. 18, 33, and 44 are shown in fig. S1.

To improve the visibility of these resonance frequencies, we applied a two-step denoising process. First, we performed a 2D Fast Fourier Transform (FFT) on the spectrogram and suppressed near-vertical features in the transformed domain. This step effectively removes common-mode instrument noise. Second, we applied Gaussian smoothing in both the time and frequency dimensions. This denoising procedure was iterated four times, each with progressively larger filter sizes, resulting in a cleaner and more continuous representation of the resonance features (fig. S1).

From the denoised spectrogram, we extracted two temporal windows (bins 150–220 and 1737–1807) to compute the mean power spectral density (PSD). Multiple peak frequencies were identified. These higher-frequency peaks typically correspond to the overtone during both early and late periods, although in some channels, the fundamental mode exhibits higher PSD (fig. S1). Notably, the frequency ratio between the overtone and the fundamental mode remains consistently close to 2.0 over time, in line with expectations for normal mode behavior.

4. Turning seismic PSD into high-resolution rain gauges

High-frequency seismic signals can serve as proxies for rainfall timing and intensity. This relationship was demonstrated by (25), who reported a Pearson correlation coefficient exceeding 0.9 between 4-minute seismic signal amplitude and precipitation rate, using a seismometer buried at a depth of 30 cm. We adopted a similar seismological processing approach to explore the correlation between rainfall and DAS amplitudes recorded at the surface.

First, we computed the power spectral density (PSD) of the DAS data in 1-minute windows. We integrated the amplitude over the 80–140 Hz frequency band to obtain the band-limited mean-square amplitude (fig. S2b). This yielded a time series of high-frequency energy at 1-minute resolution. Next, we integrated the mean-square amplitudes over 60-minute intervals to obtain hourly high-frequency energy estimates at the soil surface (fig. S2d). The resulting time series strongly correlates with hourly rain gauge measurements, confirming that the 80–140 Hz mean-square amplitudes are effective proxies for precipitation rates.

To use these amplitudes in hydrological modeling, we scaled them using the average ratio between the 1-hour integrated PSD and the corresponding rainfall rates. The scaled mean-square amplitudes were then interpreted as precipitation rate inputs for the model (fig. S2c).

Hydrological and lithological modeling of the four main rainfall events (P1–P4 in Fig. 1)— occurring at 22:02 on March 17, and at 15:14, 16:32, and 19:13 on March 18—revealed that these events significantly increased water saturation in the topsoil, resulting in decreased shear wave velocity (v_S). In contrast, modeling smaller PSD peaks preceding P1 and P2 as light rainfall led to velocity variations inconsistent with the observed changes. These weaker signals exhibit greater spatial variability (fig. S2a) and are thus interpreted as non-rainfall activities, potentially caused by different source terms, including combinations of wind, thunder, and anthropogenic activities. Spatial variation of the PSD may be partially associated with cable-soil coupling. The rainfall-related PSD remained within acceptable limits, showing lower spatial variability than non-rainfall PSD (fig. S2, A and B). Overall, the proxies of P1-P4 show relatively low spatial variability compared to other noise sources in the frequency range of 80-140 Hz.

5. Ambient noise monitoring

5.1 Time-lapse passive seismic interferometry

Time-lapse passive seismic interferometry is a method used to detect temporal changes in seismic velocity by analyzing phase shifts in the coda waves of seismic signals recorded at the same station pairs and originating from the same source conditions. These repeated waveforms are reconstructed by cross-correlating ambient seismic wavefields recorded at fixed locations. The underlying assumption of passive seismic interferometry is that, in a diffuse wavefield, coherent signals between stations can be extracted from ambient noise (27, 51, 52). In this framework, coda waves—arising from multiple scattering—represent the diffusive portion of the wavefield as it interacts with subsurface heterogeneities (53, 54). Owing to their extended propagation paths, coda waves are particularly sensitive to medium changes and relatively insensitive to variations in the noise source (55).

We applied this method to DAS data collected over 51 channels along a linear segment during a 40-hour observation period. To reduce data volume, we selected one out of every five consecutive 1-minute files, resulting in 482 DAS recordings at 5-minute intervals. The data were then processed as follows. First, each 1-minute segment was divided into four 15-second subsegments (0–15 s, 15–30 s, 30–45 s, and 45–60 s) to facilitate stacking and improve the stability of the auto-correlation function (ACF), with peak correlation coefficient measured in the coda around 80%. Second, we applied one-bit normalization and a 50-point moving average smoothing to each time-series segment. Third, we transformed the normalized data into the frequency domain and applied an additional 50-point smoothing filter to the spectra.

Passive seismic interferometry was then performed by computing the ACFs in the frequency domain. The ACFs were subsequently transformed back into the time domain and downsampled to 500 Hz. The auto-correlation window was set to span lag times from –5 to +5 seconds. The four ACFs corresponding to each 15-second sub-segment were stacked, and this process was repeated for all 482 recordings.

To evaluate frequency sensitivity, we repeated the interferometric analysis over multiple frequency bands ranging from I–2I Hz, where I = 1, 2, 3, …, 10, 15, 20, 30, 40 Hz. Clear and coherent phase evolution was observed within the 15–30 Hz, 20–40 Hz, and 30–60 Hz bands. Based on these results, we applied a bandpass filter from 15–60 Hz to the raw data and repeated the full analysis. The full result is shown in fig. S4. Examples of the 40-hour ACF evolution at Ch. 18, 33 and 44 are shown in fig. S5.

### 5.2 Temporal stretching

Assuming homogeneous velocity changes within the medium, average seismic velocity variations ($dv/v$) can be inferred from relative time shifts (dt/t) in coda waves (27):

$$\frac{dt}{t} = -\frac{dv}{v}. \tag{S1}$$

To estimate the noise correlation functions—in this case, auto-correlation functions (ACFs) across channels—we employed the core functionality of the NoisePy4DAS package (https://github.com/Denolle-Lab/FarmDAS/tree/main/utils/noisepy4das_processing). The ACFs were bandpass-filtered in the 15–60 Hz range to isolate the frequency band with the most coherent coda waveforms. We then applied the stretching method proposed by (32) to extract $dv/v$ time series for each channel. This method involves stretching or compressing the current ACF, $ACF^{cur}(t)$ to maximize its correlation with a reference ACF, $AC^{ref}(t)$. The correlation coefficient (CC) is calculated as:

$$CC(\epsilon) = \frac{\int_{t2}^{t1} AC^{cur}\big(t(1+\epsilon)\big)\, AC^{ref}(t)dt}{\sqrt{\int_{t2}^{t1}[AC^{cur}(t)]^2\, dt \int_{t2}^{t1}[AC^{ref}(t)]^2\, dt}}. \tag{S2}$$

Here, $\epsilon$ is the stretching factor, and $t1$ and $t2$ define the time window of the coda wave. We performed a grid search for $\epsilon$ in the range of -50% to +200%, with a step size of 1%. This searching range is informed by visual inspection with the average ACF taken as reference. Then, the $dv/v$ is determined by $\epsilon - 1$ with the maximum CC. (26) demonstrated that coda waves are predominantly composed of S-waves, and that $dv/v$ estimates are approximately 90% sensitive to changes in $v_S$. Accordingly, we interpret $dv/v$ as representing $dv_s/v_s$ in this study.

To ensure robustness in the velocity change estimates and reduce artifacts such as cycle skipping, we applied three iterations of denoising and optimal stretching estimation. The denoising followed the same procedure used for the spectrogram analysis: removal of vertical textures in the 2D Fourier domain (as we extract variations along the horizontal axis), followed by Gaussian smoothing in the $\epsilon$-t space. At each time point, we identified the optimal stretching factor that maximized CC. However, in periods of rapid velocity variation, temporal gaps in the measurement appeared. To address this, we used piecewise polynomial fitting to interpolate and smooth the $dv/v$ time series (fig. S6). The three-pass approach included: (1) an initial estimation of first-order velocity trends and major jumps, (2) refinement of rapid variations in response to environmental changes (e.g., rainfalls), and (3) validation that the combined signal captured the primary structure.

## 6. Lithological Model

To investigate the relationship between short-term soil moisture variations and seismic velocity, we adapted the theoretical framework of (11) by incorporating the dynamic capillary effects described by (30). In this work, a typical soil model is conceptualized as a three-phase system comprising the solid mineral matrix, the porous space, and the saturating fluids. For relatively coherent soils where the solid frame remains mechanically stable over short timescales, we assume that the solid matrix and porosity remain constant. Under this assumption, variations in the saturating fluid—specifically the degree of saturation and its distribution within the pore network—lead to changes in the soil's effective elastic moduli. Since the shear wave velocity ($v_S$) is directly dependent on these elastic properties, changes in fluid saturation can be linked to measurable variations in $v_S$. Specifically, shear velocity is given by:

$$v_S = \sqrt{\frac{\mu_{eff}}{\rho}}, \qquad (S3)$$

where $\mu_{eff}$ is the effective shear modulus and $\rho$ is the effective bulk density, respectively. The density is computed as:

$$\rho = (1 - \phi)\rho_s + \phi[S_w\rho_w + (1 - S_w)\rho_a]. \qquad (S4)$$

Here, $\phi$ is the porosity (measured at 10 cm, 20 cm, and 30 cm depths for each plot), $\rho_w$, $\rho_a$ and $\rho_s$ are the densities of water, air, and soil matrix, respectively, and $S_w$ is the pore water saturation. For modeling, we use the porosity values at 10 cm depth.

To estimate the effective shear modulus $\mu_{eff}$ (shear modulus of the soil's frame), we apply the Hertz-Mindlin contact theory (29), assuming the soil behaves as a granular medium. In this model, the soil's frame is idealized as a random packing of identical spherical grains with porosity $\phi$ and an average number of contacts $N$. Previous studies have shown that $\mu_{eff}$ depends on the effective confining pressure $P_e$, the grain Poisson's ratio $\nu_s$, and the interparticle contact behavior, notably the fraction of non-slipping grains f (11, 56–59). The grain Poisson's ratio $\nu$ is defined as,

$$\nu = \frac{3K_s - 2\mu_s}{2(3K_s + \mu_s)},\qquad (S5)$$

where $K_s$ and $\mu_s$ are the bulk and shear moduli of the soil grains, respectively.

The effective confining pressure $P_e$ incorporates gravitational and capillary effects and is given by:

$$P_e = \rho_b gz - \rho_a gz(1 - S_{we}) - \rho_w gz S_{we} - \tau(S_w)\frac{\partial S_w}{\partial t}. \qquad (S6)$$

Here, $g$ is the gravitational acceleration, $z$ is depth, and $S_{we} = (S_w - S_{wr})/(1 - S_{wr})$ is the effective saturation accounts for the static capillary suction stress with a residual water saturation $S_{wr}$. Residual water saturation corresponds to the irreducible saturation level during the drainage process, below which water content no longer decreases when decreasing water-phase boundary pressure. We use $S_{wr} = 0.25$ for clay and $S_{wr} = 0.15$ for sand, following (11).

The final term in equation (S6) represents dynamic capillary effects—capillary suction changes during wetting or drying cycles. Following the dynamic capillarity theory (30, 31, 60–64) this effect is modeled as the product of a dynamic coefficient $\tau$ and the time derivative of saturation $\frac{\partial S_w}{\partial t}$, leading to hysteretic evolution of effective shear modulus (fig. S7). (30) developed an experiment with fine-grain soils with porosity 0.45, finding two constitutive relationships for the wetting and drying cycles. We adopt different expressions of $\tau$ for wetting and drying cycles and extrapolate the relationship for $S_w < 0.4$. For drying cycles, we took the log-linear dynamic coefficient ($\tau = 5 \times 10^5 e^{S_w^5}$). For drying cycles, we took a log-linear relationship ($\tau = 1.5 \times 10^{10} e^{-9 \times S_w}$). This combination of dynamic coefficients results in pronounced hysteresis in the $\nu_S$ model (fig. S7), capturing the distinct responses of soil stiffness during wetting and drying cycles. When the soil is fully saturated, the model simplifies to the classic effective stress formulation $P_{overburden} - P_{hydrostatic}$.

We demonstrate the importance of incorporating the dynamic capillary effect by comparing three lithological models for the high-disturbance soil in Fig. 2b: one with dynamic capillary stress, one with static capillary stress, and one without any capillary effect. The model, including the dynamic capillary term, accurately captures both the absolute values of shear wave velocity and its rate of change—showing accelerated velocity increases during evaporation and pronounced drops during precipitation events. In contrast, the static capillary model underestimates the velocity increase during evaporation by approximately 50% and slightly slows the modeled drainage rate. The model without any capillary stress produces significantly smaller velocity variations during both wetting and drying, indicating that capillary effects are essential for reproducing observed velocity dynamics. In contrast, Fig. 2c shows that the capillary effect is negligible for Ch. 33, which experienced minimal disturbance.

Given the effective pressure $P_e$, and Poisson's ratio $\nu$ and the soil grain's shear modulus $\mu_s$, we can compute the effective shear modulus $\mu_{eff}$ of the soil's frame:

$$\mu_{eff} = \frac{2 + 3f - (1 + 3f)\nu}{5(2 - \nu)} \left[\frac{3N^2(1 - \phi)^2 \mu_s^2}{2\pi^2(1 - \nu)^2} P_e\right]^{\frac{1}{3}}. \qquad (S7)$$

Here, $f$ is the fraction of non-slipping particles, and $N$ is the average number of grain-to-grain contacts per particle. In this study, we explore a range of f values (0.01–0.50) as a function of water saturation and adopt $N = 6$ and 8 for sand and clay, respectively, following (11). The solid grain properties are also based on their study. Based on the formulation above, the shear-wave velocity increase is proportional to saturation reduction, but that the slope depends on the style of capillary response (fig. S12). To estimate the effective pressure $P_e$, we solve for water saturation in the pore space using a water budget model, as described in the following section.

### 7. Hydrological Model

We focus on the short-term water flow dynamics within the topsoil (~10 cm), where the water content responds rapidly—seconds to minutes—to meteorological drivers such as solar radiation, vapor pressure, and precipitation. To capture this, we adopt a first-order water balance model assuming horizontally homogeneous and vertically averaged soil properties, following an approach similar to (28). In this model, water is recharged by precipitation $P(t)$, and lost via evapotranspiration $ET(t)$, and vertical drainage $Q(t)$. Considering a control volume extending from the surface to a depth z, the water budget is expressed as:

$$z \, \phi \, \frac{dS_w}{dt} = P(t) - Q(t) - ET(t, D, S_w). \qquad (S8)$$

Here, the evapotranspiration term reflects bare-soil evaporation (no vegetation), modeled as: $ET(t, S_w, DI) = DI \, S_w(t) \, K_C \, ET_0$, where $DI$ is a disturbance index specific to each plot (representing tillage and compaction effects), $K_C = 1$ for bare soil (33), and $ET_0$ is the reference evapotranspiration every 1 minute computed using the Penman-Monteith equation (33),

$$ET_0 = \frac{0.408\Delta(R_n - G) + \gamma \dfrac{37}{60(T + 273)} u(e_s - e_a)}{\Delta + \gamma(1 + 0.34u)}. \qquad (S9)$$

In this formulation: $R_n$ is net radiation at the soil surface (MJ m-2 min-1), $G$ is soil heat flux density (in MJ m-2 min-1), $T$ is air temperature (in °C) from the Harper Adams weather station, $u$ is wind speed (in m/s) from the Newport weather station , $e_s$ is saturation vapor pressure calculated as $e_s = 0.6108 e^{\frac{17.27T}{T+237.3}}$, $e_a$ is actual vapor pressure given by $RH \times e_s$, $\Delta$ is the slope of saturation vapor pressure-temperature curve at temperature $T$ given by $4098e_s/(T + 237.3)^2$, and $\gamma = $ is psychrometric constant ($0.665 \times 10^{-3} \times 101.325$ kPa °C-1). We model the net radiation based on weather conditions, the angle of the sun, and the time of day, following the same formula used in equation 28 of (33). When the soil can supply water fast enough, $ET_0$ is determined solely by these meteorological conditions.

The reference evapotranspiration $ET_0$ represents the rate of evapotranspiration from a standardized soil surface, well-watered and vegetated uniformly by 0.12-m tall reference crop with an albedo of 0.23 and surface resistance of 70 s/m. In practice, $ET_0$ is primarily driven by solar radiation and, to a lesser extent, air temperature, with wind and vapor pressure deficit modulating the rate of surface drying. We model $ET_0$ using the weather data collected from Newport and Harper Adams weather stations. The result is shown in Fig. 1F.

Precipitation $P(t)$ is the only water input, estimated using high-frequency PSD. Lateral runoff is neglected due to the flatness of the study site. For vertical drainage, we adopt a simplified Darcy baseflow model (32, 65, 66).

$$W = P(t) - Q(t, D) \qquad (S10)$$

Here, the vertical drainage $Q$ modeled as a convolution of rainfall with an exponential kernel, approximating a drainage timescale of ~70 hours and a nominal infiltration depth of 0.7– 1 m, consistent with a hydraulic conductivity of ~10 mm/h. Specifically, we apply a smoothing kernel $e^{-\frac{t}{3200}}$ to model the exponentially decreasing contribution of prior rainfall to current soil water content. Combining ET and W, we simulated the temporal evolution of water saturation $S_w(t)$ in the pore space (Fig. 2a and fig. S8). This saturation is then fed into our lithological model to predict shear wave velocity $v_s$, as shown in Fig. 2b, 2c and fig. S8, capturing both drying and wetting cycles with physically grounded hydrological controls. The model $dv/v$ of all channels is shown in fig. S9.

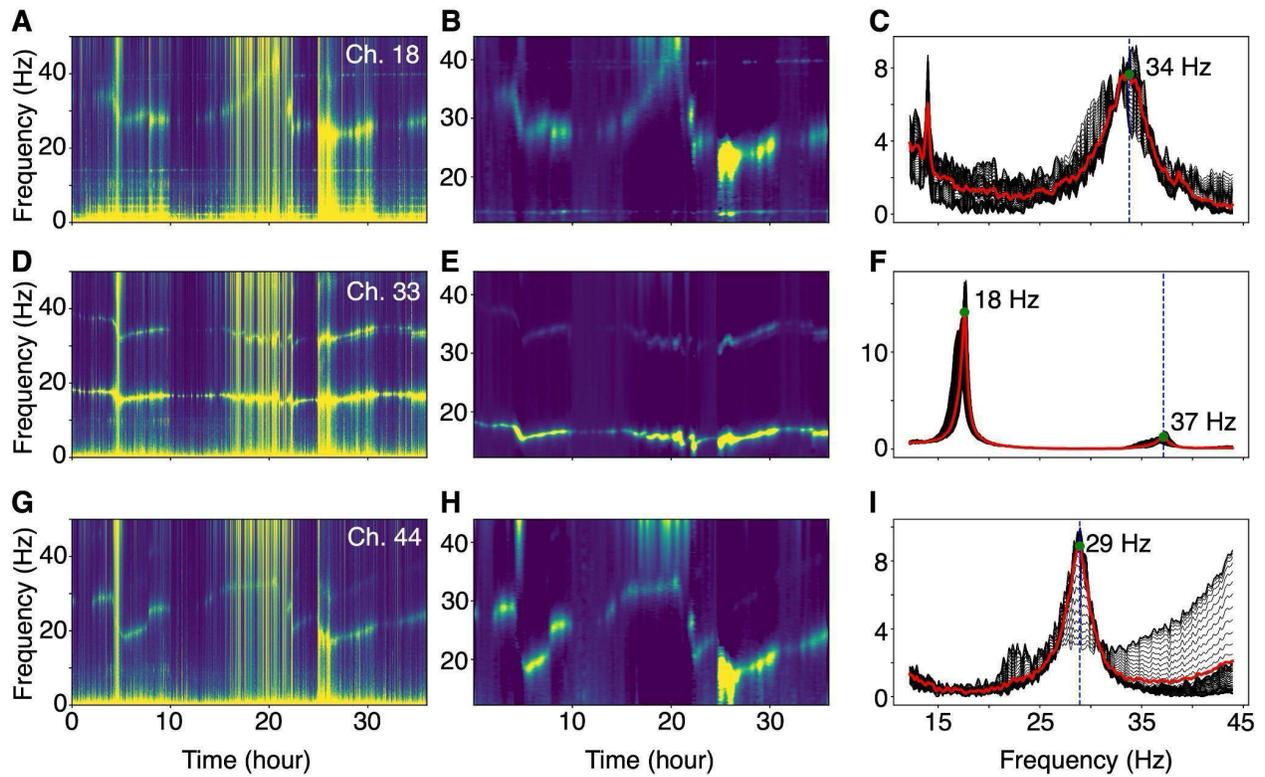

**Fig. S1. Resonance modes and denoising.**
(**A**) Raw spectrogram at Ch. 18. (**B**) Denoised and spectrogram at Ch. 18, truncated for 11-45 Hz. (**C**) PSDs of the beginning 150- 220th time bins (black) and the mean PSD (red) with peak PSDs denoted by green dots. (**D-F**) Similar to A-C, but for Ch. 33. (**G-I**) Similar to A-C, but for Ch. 44.

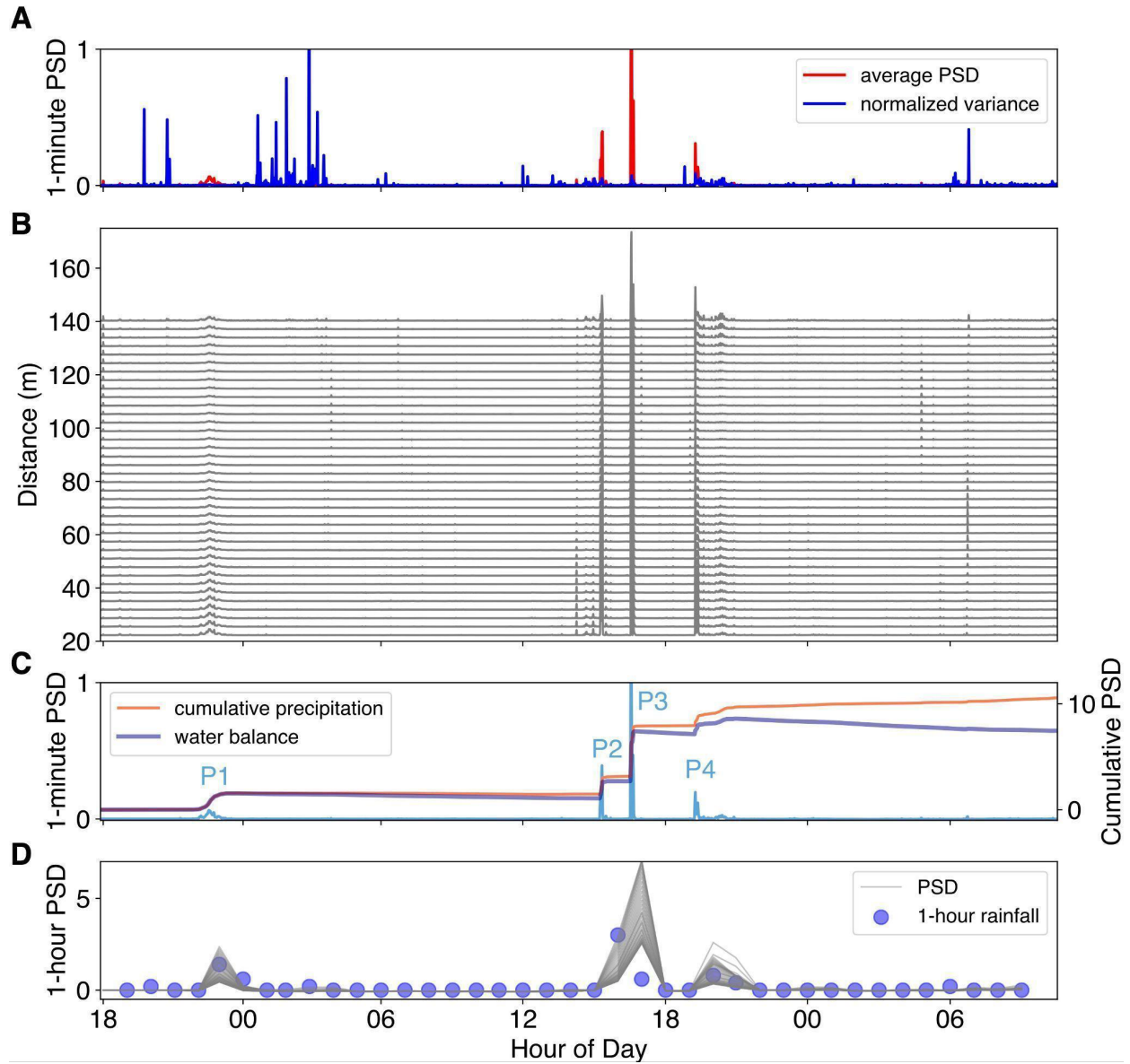

**Fig. S2. Rain gauge by DAS data.**
(**A**) Space-averaged 1-minute PSD (red) and its variance (blue) normalized by its amplitude. (**B**) 1-minute PSD sorted by distance. (**C**) PSD with exponential decay (lightblue) representing stable drainage, and the cumulative effect (orange and navy). (**D**) PSD summed for every hour (gray lines), compared with the hourly rainfall recorded by the onsite rain gauge (blue dots).

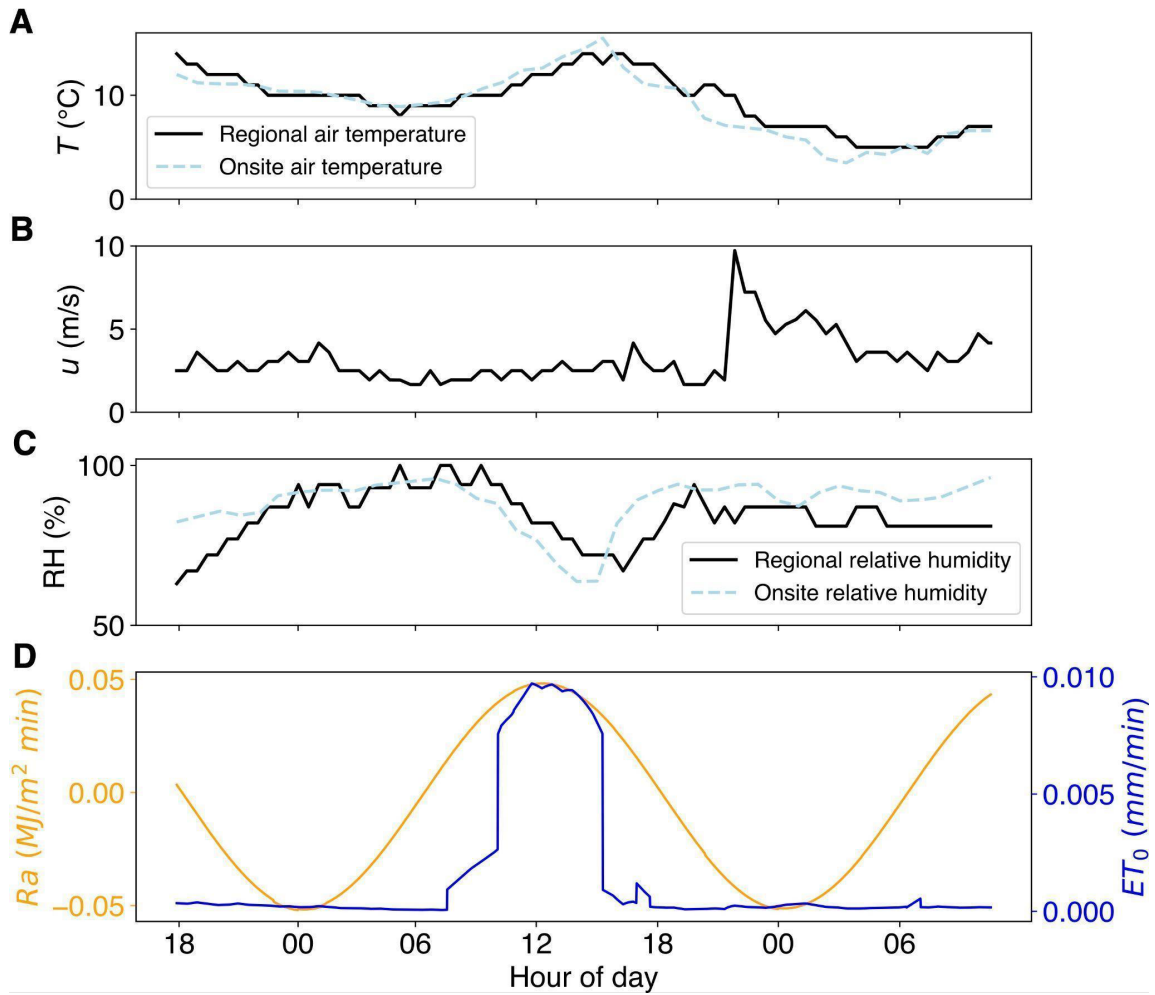

**Fig. S3. Weather data and evapotranspiration.**

(**A**) Temperature was measured by the Newport regional weather station (black) and the onsite thermometer at Harper Adams University (light blue) (**B**). Wind speed was recorded at the Newport station. (**C**) Relative air humidity on Newport station (black) and the onsite station (light blue). (**D**) Extraterrestrial radiation for the research farm (orange) and the reference ET derived from the weather data.

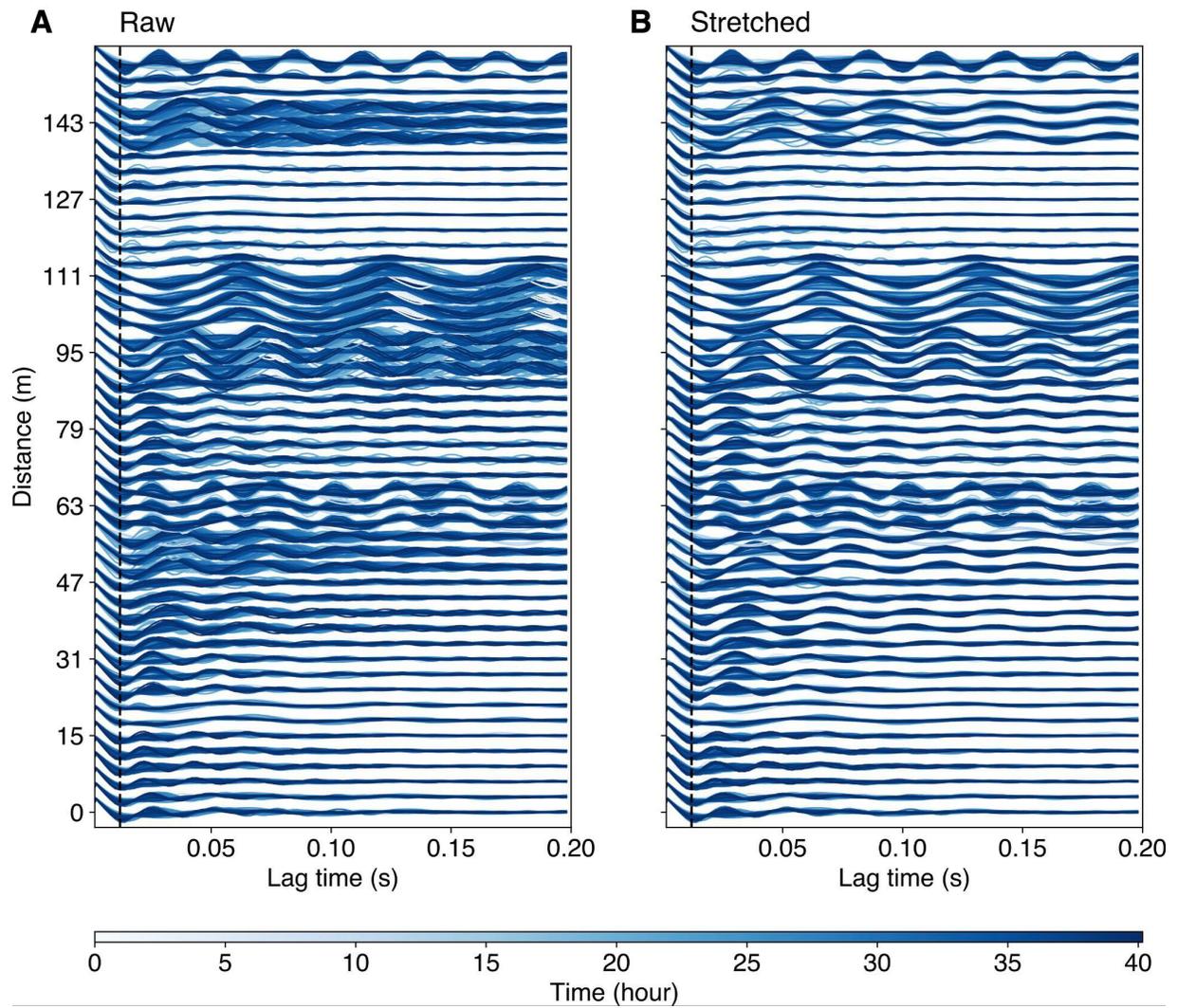

**Fig. S4. Auto-correlation functions (ACFs) of all channels.**

(**A**) Raw ACFs, with the x-axis showing the lag time and the y-axis showing the optical distance. The ACFs are color-coded by clock time. The starting time of the coda wave for $dv/v$ estimation is indicated by a dashed line (t=0.012 s). (**B**) The stretched ACFs, presented in the same fashion as A.

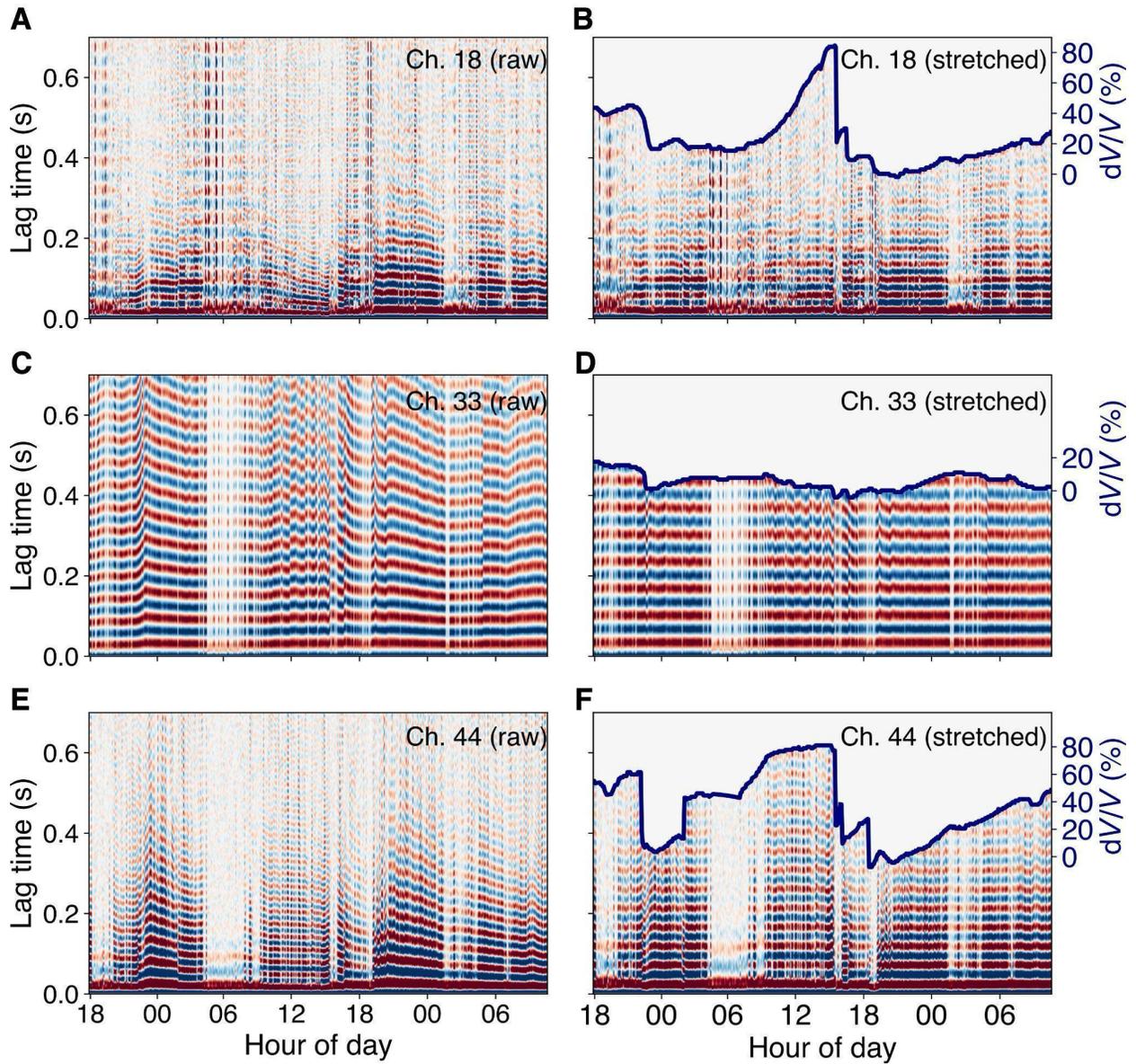

**Fig. S5. Auto-correlation functions (ACFs) of Ch. 18 and 33.**

(**A**) ACFs over 40 hours at Ch. 18, with positive amplitudes indicated in blue and negative amplitudes in red. (**B**) Stretched ACFs of Ch. 18, with the estimated $dv/v$ shown in navy blue. (**C**, **D**) Similar to A and B but for Ch. 33. (**E**, **F**) Similar to A and B but for Ch. 44.

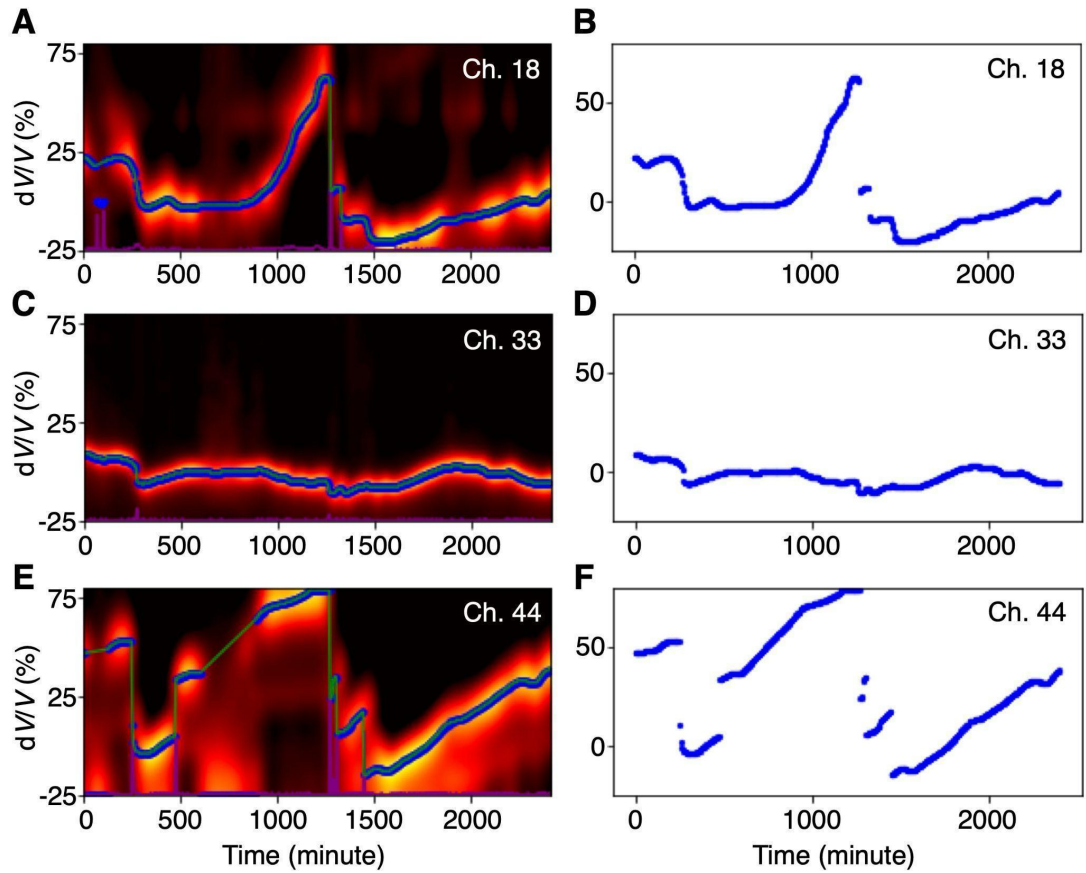

**Fig. S6. Velocity changes from stretching and denoising.**

(**A**) Denoised $dv/v$ $t$ map at Ch. 18, with the light color indicating the higher Pearson coefficient between the ACF at each time bin and the reference ACF. The peak Pearson coefficients are picked as blue dots. (**B**) Estimated velocity change of Ch. 18. (**C**, **D**) Similar to A and B, but for Ch. 33. (**E**, **F**) Similar to A and B but for Ch. 44.

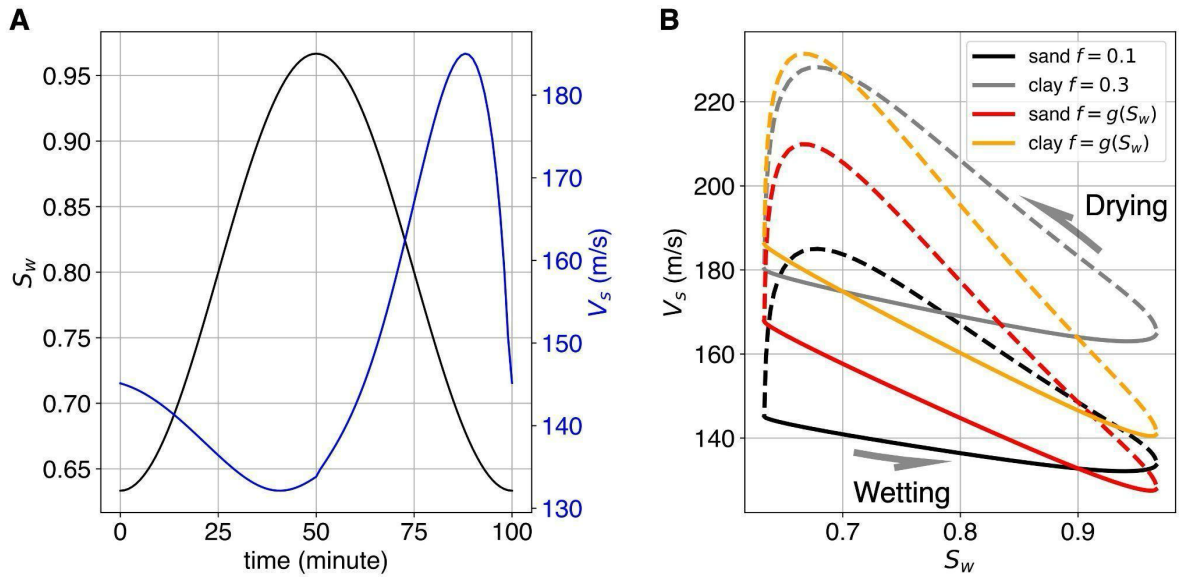

**Fig. S7: Hysteresis of the wetting-drying cycles.**

(**A**) A cosine-shaped water saturation change ($S_w$: black) represents a 50-minute wetting process followed by a 50-minute drying process. The velocity evolution is calculated ($v_S$: blue) with dynamic capillary effect. (**B**) The hysteretic behavior of velocity for sand- and clay-type soils, respectively, with constant or varying non-slipping fractions f.

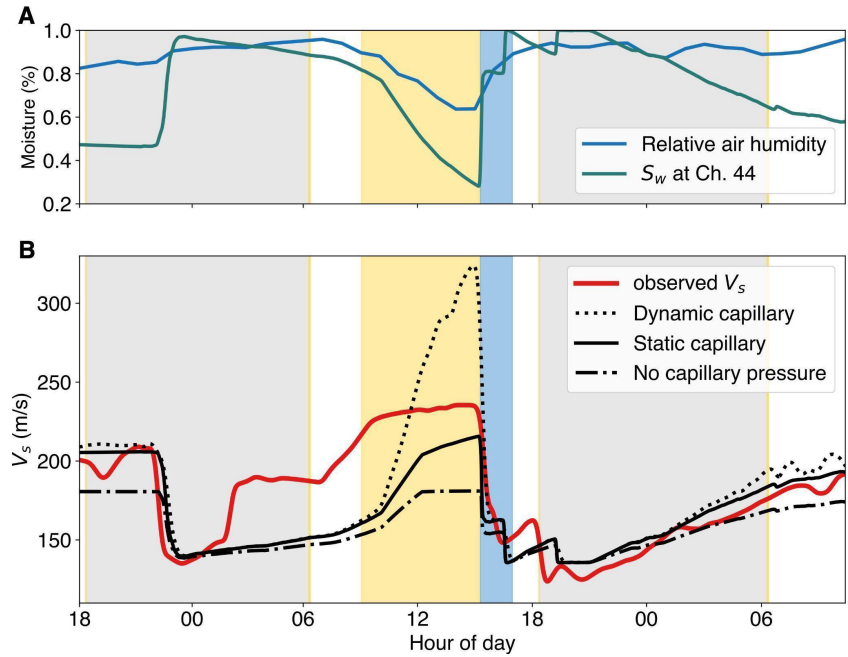

**Fig. S8: Hydrological and lithological responses at a medium-disturbance plot.**

(**A**) Hydrological response from the modelled saturation ($S_w$), compared with the relative air humidity. (**B**) Observed velocity change (red) compared with lithological models with dynamic capillary stress, static capillary stress, and no capillary stress.

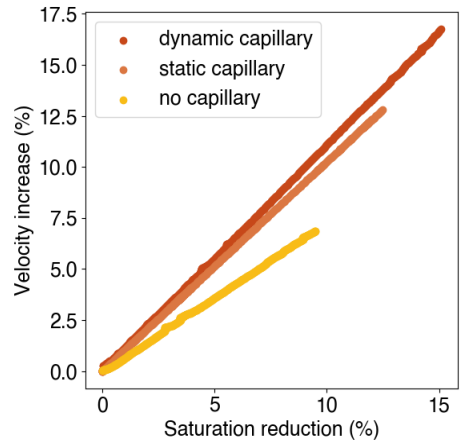

**Fig. S9. Lithological versus hydrological responses.**

Relative velocity changes in relation with relative hydrological variability during the drainage process, color-coded for dynamic capillary, static capillary and no capillary regimes.

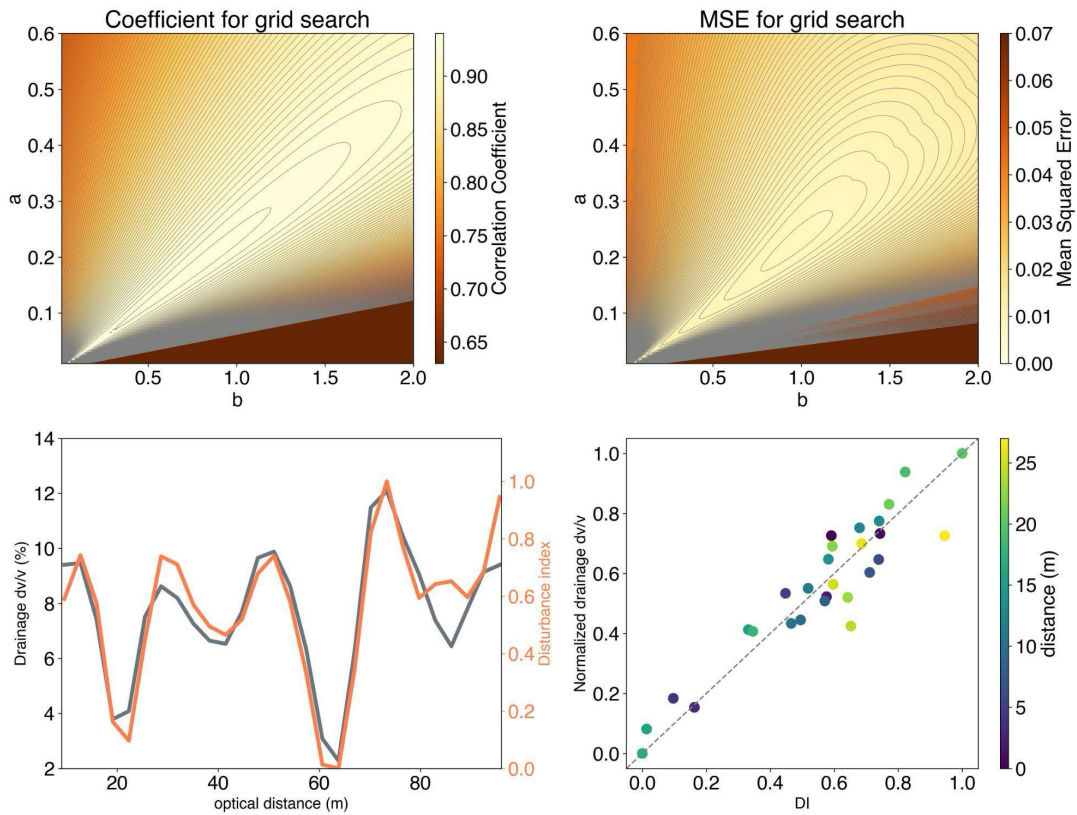

**Fig. S10: Disturbance index (DI).**

(**A**) Grid-search of the exponents of tillage depth and compaction pressure based on spatial cross-correlation. (**B**) Grid-search based on mean-squared-error (MSE). (**C**) Velocity change compared with the DI in space. (**D**) Similar to (**C**) but each plot is represented as s scatter point.

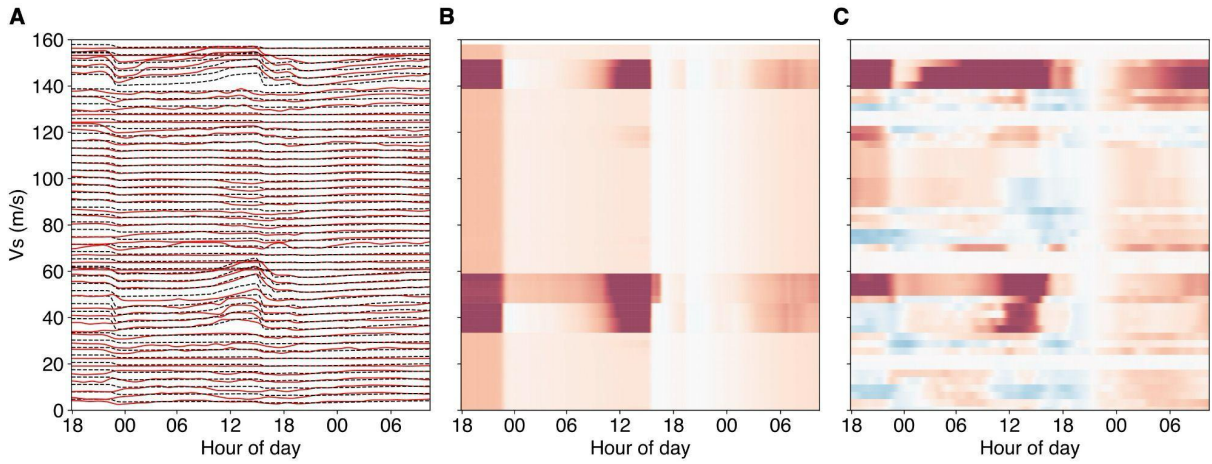

**Fig. S11: Modelling velocity variations in space.**

(**A**) Observed $dv/v$ along the cable distance (red lines) and the modelled $dv/v$ (black dashed lines). (**B**) The modelled $dv/v$ in space and time, colorcoded by amplitude. (**C**) The observed $dv/v$ in space and time, color-coded by amplitude, with red indicating higher $v_S$.

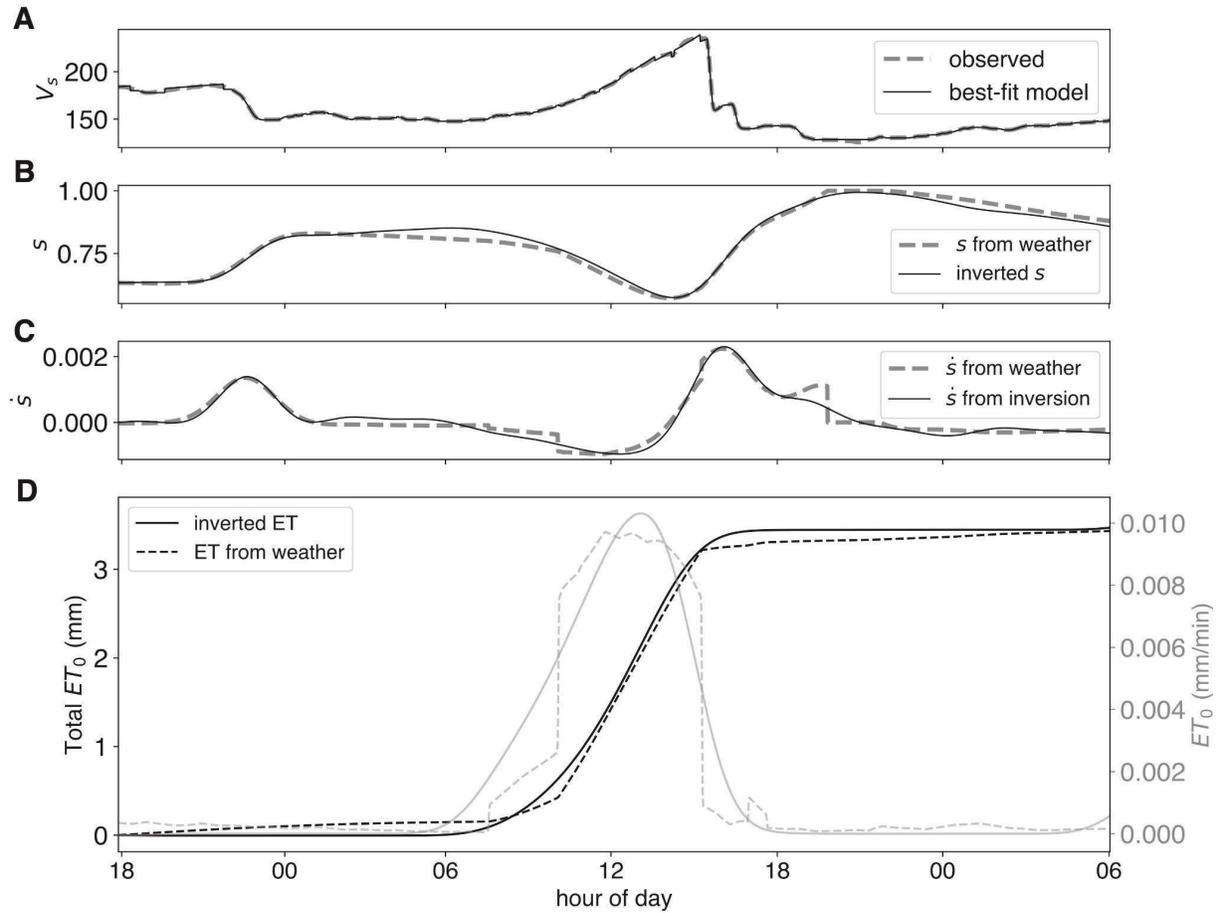

**Fig. S12: Seismically estimated evapotranspiration (ET).**

(**A**) The observed $v_S$ (dashed) and the synthetic $v_S$ from the inverted $S_w$ (solid). (**B**) The $S_w$ derived from weather data (dashed) and inverted $S_w$ (solid). (**C**) Time derivatives of $S_w$ calculated from weather data (dashed) and from seismic inversion (solid). (**D**) $ET_0$ inverted from $S_w$ (solid) and derived from weather data (dashed). The 1-minute $ET_0$ rate is shown in gray, and the cumulative $ET_0$ is shown in black.